# 融智学视域下对外汉语教学新模式

邹晓辉[1, 2*],柯丽君[2],邹顺鹏[3]

[1]北京大学教师教育发展中心(跨学科知识建模课题组),北京 100871
[2]珠海横琴塞尔科技有限公司中美塞尔研究中心,广东珠海 519000
[3]英国布达佩斯国际学院鹏哥教学中心(对华招生部),匈牙利布达佩斯 1025

邮箱
zouxiaohui@pku.org.cn , 949309225@qq.com , robertzou.sp@hotmail.com

摘要:本研究旨在介绍融智学视域下对外汉语教学新模式。其特点是:在方法聚焦于先解释后翻译的蝴蝶模型,突出双语思维训练新方法,一方面,应用汉字汉语新理论,言和语的关系理论,语言科学的前瞻性研究成果;另一方面,应用对外汉语教学新模式,AI赋能教与学,教育科学的前瞻性研究成果,充分体现融智学视域下对外汉语教学新模式的一系列特点。其有益效果是:不仅改变了旧的语言观和旧的教育观尤其是旧的对外汉语教学观,而且还改变了旧的人机交互观。其意义在于:从融智学视域,明确提出了语言、知识、教育、教学等一系列大跨界的学问及其双语思维训练新方法和新课题,尤其是在Chat GPT对人类的学习能力乃至创造能力的挑战面前,现有的语言知识教育教学观念等都已经很落后了,旧的汉语观和旧的教育观以及旧的对外汉语教学观都面临着一系列的颠覆性创新的挑战,如何在适应当中求变革?本研究做了一系列创新尝试,希望惠及学界同仁及广大师生。

摘要附图:图21 (巧用GXPS及其调用产生的Chat GPS做)与Chat GPT呼应的机人间接对话和人机直接对话。

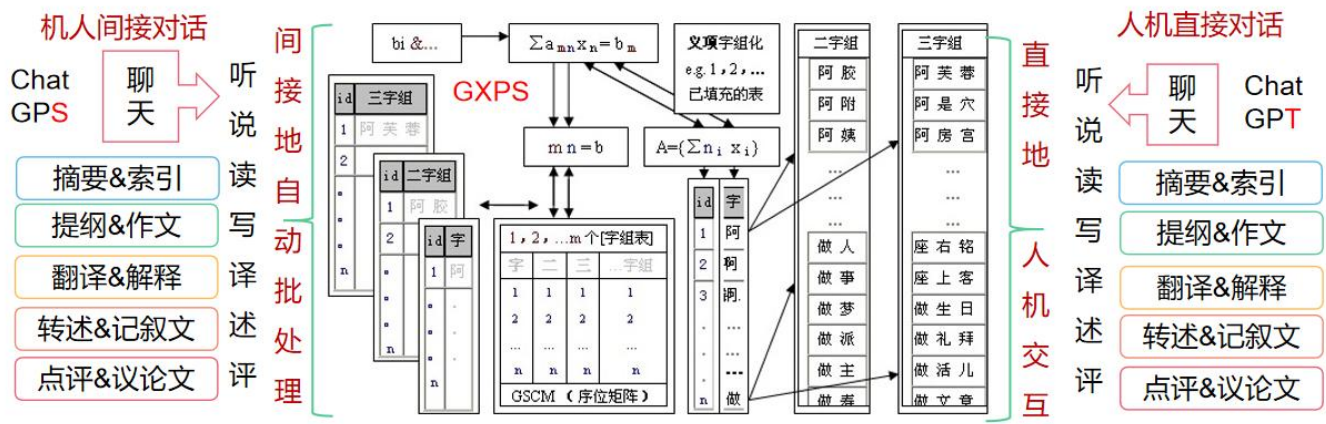

关键词:对外汉语教学新模式,融智学视域,汉字汉语新理论,AI赋能教与学,先解释后翻译的蝴蝶模型,言和语的关系,双语思维训练新方法,习惯用语的刻意练习

---

# A New Mode of Teaching Chinese as a Foreign Language from the Perspective of Smart System Studied by Using Rongzhixue

Xiaohui Zou[1,2 *], Lijun Ke[2], Shunpeng Zou[3]

[1]Teacher Education Development Center (Interdisciplinary Knowledge Modeling Research Group), Peking University, Beijing 100871, China
[2]Sino-US Searle Research Center, Zhuhai Hengqin Searle Technology Co., Ltd. Hengqin, Zhuhai 519000, China
[3]Peng Ge Learning Center (Admissions Department for China), Budapest British International Academy, Budapest 1025, Hungary

Email address
zouxiaohui@pku.org.cn , 949309225@qq.com , robertzou.sp@hotmail.com

Abstract: The purpose of this study is to introduce a new model of teaching Chinese as a foreign language from the perspective of integrating wisdom. Its characteristics are as follows: focusing on the butterfly model of interpretation before translation, highlighting the new method of bilingual thinking training, on the one hand, applying the new theory of Chinese characters, the theory of the relationship between language and speech, and the forward-looking research results of language science; On the other hand, the application of the new model of teaching Chinese as a foreign language, AI empowering teaching and learning, and the forward-looking research results of educational science fully reflect a series of characteristics of the new model of teaching Chinese as a foreign language from the perspective of integrating wisdom. Its beneficial effects are: not only the old view of language and education, especially the old view of teaching Chinese as a foreign language, but also the old view of human- computer interaction. Its significance lies in that a series of great cross-border Rongzhixue such as language, knowledge, education and teaching, as well as new methods and new topics of bilingual thinking training are clearly put forward from the perspective of integrating wisdom. Especially in the face of the challenge of Chat GPT to human learning ability and even creativity, the existing concepts of language knowledge education and teaching are already very backward. The old concepts of Chinese language education, and teaching Chinese as a foreign language are all facing a series of subversive innovation challenges. How to seek changes in adaptation? This study has made a series of innovative attempts, hoping to benefit academic colleagues, teachers and students.

## 1. 引言

本研究旨在从融智学大跨界视域论对外汉语教学。其背景涉及语言科学研究的最新成果、双语记忆模型、第二语言习得理论、中介语假说、七遍通方法以及对外汉语教学的基本原则和专家经验。其中"言和语的关系：自然语言的形式化理解"以及"言识软硬教管学用文化基因系统工程"涉及语言科学和教育科学的最新成果。

本研究旨在介绍融智学视域下对外汉语教学新模式。其特点是：聚焦于先解释后翻译的蝴蝶模型，突出双语思维训练新方法，一方面，应用汉字汉语新理论，言和语的关系理论，语言科学的前瞻性研究成果；另一方面，应用对外汉语教学新模式，AI赋能教与学，教育科学的前瞻性研究成果，充分体现融智学视域下对外汉语教学新模式的一系列特点。

本文破例提供了摘要附图（借鉴专利说明书摘要的方式）：图21 （巧用GXPS及其调用产生的Chat GPS做）与Chat GPT呼应的机人间接对话和人机直接对话。这是本研究的目的、方案和有益效果的集中呈现。正文详论。

## 2. 正文

什么是融智学视域下对外汉语教学新模式？突破点在哪里？以拥有的三类双语支持做对外汉语教学资源库的方式方法，线上线下结合满足师生们与Chat GPT首先用英语对话，进而用英汉双语对话，最终用地道的汉语对话。在这个过程中，不仅加速了对外汉语教学的进度，而且还大大提升了对外汉语教学的质量，并且是可持续的，也是投入产出效益暨社会文教经济综合效益的性价比最优的。

### 2.1. 第二语言习得理论

表1 第二语言习得理论：克拉申（D. Krashen）五个假设[1]。

| 表题 | 第二语言习得理论：克拉申（D. Krashen）五个假设 | |
|---|---|---|
| | 汉语 | 英语 |
| 1 | 习得-学习差异假设 | Acquisition-Learning Hypothesis, ALH |
| 2 | 自然顺序假设 | Natural Order Hypothesis, NOH |
| 3 | 监检假设 | Monitor Hypothesis, MH |
| 4 | 输入假设 | Input Hypothesis, IH |
| 5 | 情感过滤假设 | Affective Filter Hypothesis, AFH |

由表1可见，克拉申五个假设，通过对第二语言习得过程的分析，系统地阐述了外语教学思想和体系。出发点及其核心是对"习得"和"学习"的区分，明确了两者的差异及其各自在习得者第二语言能力形成过程中所起的不同作用。克拉申他指出成人是通过两条截然不同的途径逐步习得第二语言能力的。第一途径是"语言习得"过程类似于儿童母语能力发展的过程，是一种无意识地、自然而然地学习第二语言的过程。第二途径是"语言学习"即通过听教师讲解语言现象和语法规则，并辅之以有意识的练习、记忆等活动，达到对所学语言的了解以及对其语法概念的"掌握"。习得潜意识的语言能力；学习有意识地掌握语言结构。克拉申强调"习得"是首要的、第一位的。但也并不排斥"学习"的作用尤其是监检作用。

自然顺序假设人们对语言结构知识习得过程实际上是按一定顺序进行的，该次序是可预测的。有些语言结构先习得，另一些语言结构后习得。克拉申指出自然顺序并不要求人们按这种顺序来制定教学大纲。实际上如果我们的目的是要习得某种语言能力，那么，就有理由不按任何语法顺序来教学。

监检假设体现出"语言习得"与"语言学习"的内在关系。鉴于两者作用各不相同，语言习得能"引导"我们讲第二语言，并直接关系到我们说话的流利程度；而语言学习只起监检或"编辑"的作用。由"习得"系统产生，经"学习"系统监检成为"真言"而说出话语。语言学习的这种监检功能，既可能在我们说话或写作之前，也可能在其后，能否充分发挥作用？有三个条件：1)语言使用者必须要有足够习练时间才能有效地选择和运用语法规则；2)语言使用者的注意力必须集中在所用语言形式上，必须考虑语言的正确性；3)语言使用者必须已具有所学语言的概念及规则等知识。在口头交谈中，人们往往没有时间去考虑斟酌语法，注重的是说话内容而不是形式，语法规则如果不是习得的，一时也用不上。因此，在口头交流中，如果一方过多地使用语法监检，不时地纠正自己讲话中的语法错误，说话就会不顺畅，对方也会失去继续交谈耐心，更达不到交流目的。但是在事先准备的正式发言和写作中，语法的使用能提高语言的准确性，为演讲或文章增添色彩。克拉申区别了三种不同的监检使用类型：第一种是使用得比较成功的。这种人在口头使用语言时常有失言，但经人指出后能够自己改正；在写作时由于较注重语言的形式，很少会出现错误。第二种是使用过度的人。这种人对语言规则懂得很多，却不敢作口头表达，书面语一般都较准确。第三种是使用不足的人。这种人能口头表达，但错误很多，不能自己改正。

输入假设是克拉申理论的核心部分。这一假设表明了他对第二语言习得者如何接受并吸收语言材料这一过程的实质性认识。以前的外语教学由于受结构主义语言学的影响，大力提倡先学句型即句子结构，然后再将这些学到的句型用于交际中加以练习。以为只有这样才有可能培养学生流畅地说外语的能力。而克拉申则说，只有当习得者接触到"可理解的语言输入"(comprehensive input)略高于其现有语言技能水平的第二语言输入，能把注意力集中于对意义或信息的理解而不是对形式的理解的时候，才产生习得。如果习得者现有水平为"i"，能促进他习得就是"i+1"的输入。根据克拉申的观点，这种"i+1"的输入并不需要人们故意地去提供，只要习得者能理解输入，又有足够的量时，就自动地提供了这种输入。按照输入假设，说话的流利程度是随时间的流逝自然而然地达到的，不能直接教会。

克拉申认为，理想的输入应具备以下几个特点：
1)可理解性comprehensibility。 理解输入语言的编码信息是语言习得的必要条件，不可理解的输入就只是一种噪音(It's incomprehensible)。2)既有趣又有关 (interesting and relevant)。 要使语言输入对语言的习得有利，必须对它的意义进行加工，输入的语言材料越有趣、越有关联，学习者就会在不知不觉中习得语言。3)非语法程序安排(not grammatically sequenced)。 语言习得关键就是足量的可理解的输入。如果目的是"习得"而不是"学习"，那么，按语法程序安排的教学不仅不必要而且不足取。4)要有足够的输入量。要习得一个新的语言结构，单靠几道练习或者几篇短文远远不够，它需要连续多个小时的有内容有乐趣的广泛阅读和许多的会话才能达到目的或解决问题。

情感过滤假设，有大量的适合的输入的环境并不等于学生就可以学好目的语即外语，第二语言即外语的习得的进程还受许多情感因素的影响。因此，语言输入必须通过情感过滤才有可能变成语言"吸入"(intake)。 在语言输入到达大脑的语言习得器官的过程中，过滤就是输入的语言信息必须逾越的头一道障碍。情感因素起着促进或阻碍的作用。克拉申在总结过去第二语言教学中的经验时指出，在习得第二语言或外语的过程中习得者不吸收他所听到的一切，有几个心理因素影响着习得语言的速度和质量。它们是：1)动力。学生的学习目的是否明确，直接影响其学习效果。目的明确则动力大，进步快；反之则收效甚微。2)性格。所有文献几乎都一致表明，比较自信，性格外向，乐于把自己置于不熟悉的学习环境，自我感觉良好的学生在学习中进步较快。3) 情感状态。主要指焦虑和放松即互为对照的精神状态。在较大程度上影响外界的语言输入。在第二语言或外语的学习中焦虑较少的人容易得到更多的语言输入。心情放松和感觉舒适的人在较短时间内学得更好。[2-7]

2.2. 双语记忆模型及其表征层级模型

表2 双语记忆模型及其表征层级模型。

| 表题 | 双语记忆模型及其表征层级模型 | | |
|---|---|---|---|
| | 创立者 | 英语 | 汉语 |
| 1 | Kolers, Paul A. 双语记忆模型 | Separate Store Model | 独立存储模型 |
| 2 | | Shared Store Model | 共同存储模型 |
| 3 | Potter, M. et. al. 表征层级模型 | Word Association Model | 单词联想模型 |
| 4 | | Concept Mediation Model | 概念中介模型 |
| 5 | Kroll and Stewart 修正的层级模型 | Asymmetry Model | 非对称模型 |

由表2可见，共同存储模型（Shared Store Model）和独立存储模型（Separate Store Model）[8]是Kolers, Paul A.提出的研究成果。共同存储模型认为双语者的知识和经验不依赖于获取该知识经验的语言通道，而是以某种超语言的抽象的概念形式储存的，两个语言通道输入的词汇信息在大脑中形成一个单一的语义认知表征。独立存储模型则认为经验和记忆以最初获得它们的语言被分别加以存储，因此，双语者在头脑中建立了两个相应的语言存储系统，即语言1系统和语言2系统，分别处理来自语言1和语言2的信息同时，与两种语言的词汇相联系的概念表征也不同。

Kolers对两个模型进行了检验，他要求双语被试进行语言内和跨语言联想，结果发现所有被试都用一种语言的联想大约有1/3是另一种语言中联想的翻译词，大约有2/3的翻译词对的联想是不同的。据此他认为双语者的经验和记忆是被表征在不同的系统中的，是独立存储的。

McLeod使用系列词表学习任务发现了跨语言系统的学习迁移效应，即对一种语言的单词进行学习能够促进对另一种语言中的翻译词的学习，支持共同存储模型。[9]

Potter等人以此为基础提出双语记忆表征的层级模型，包括两个假设：单词联想模型 (Word Association Model)和概念中介模型 (Concept Mediation Model)。前者是假设两种语言中的单词与概念的联系是不同的。即第一语言(L1) 中的词汇直接与概念联系，第二语言 (L2) 中的单词只能通过 L1 中的单词间接地与概念发生联系，即在进行跨语言加工时两种语言中的词汇通过词汇通路直接进行转换，不必通过概念的中介。后者则假设L1和L2中的词都

与概念有直接的联系。因此，在进行跨语言加工时，两种语言中的词汇都要通过概念进行转换。Potter等人比较了L2非常熟练和不熟练的中英双语大学生用L2翻译单词和命名图片的反应时，结果发现，L2 熟练程度不同的两组被试单词翻译和图片命名的反应时，没有显著差异，表明从L1到L2的单词翻译是概念中介的。这一结果支持概念中介模型。[10]

Kroll和Curley发现成人初学者用L2翻译单词比用L2命名图片更快，表明他们在进行翻译的时候，是通过词汇联系进行的。熟练双语者的反应结果支持概念中介模型。这意味着，随着L2熟练程度的提高，双语者的记忆表征有一个从单词联想模型发展到概念中介模型的过程。[11]

Kroll和Stewart修正的层级模型假设在双语者的记忆中词汇和概念联系都是活跃的。但是这些联系的强度随着L2的熟练程度和L1对L2的相对优势而变化。该模型综合了单词联想模型和概念中介模型的内容。强调两种语言间词汇联系的不对称性，两种语言中词汇与概念联系强度的不对称性，因此，该模型又被称为：非对称模型(The Asymmetry Model)。[12]

非对称模型得到了许多支持证据。一些研究发现 L2不熟练的成人双语被试从L1到L2的翻译，要比L2到L1的翻译慢，而且前者受到语义变量的影响，而后者不受语义变量的影响。这表明L1到L2的翻译是概念中介的，而L2到L1的翻译是词汇中介的。

使用跨语言语义启动任务的一些研究也发现L1词可对L2词产生显著的语义启动效应。L2词对L1词不能产生启动效应。即使产生启动效应也显著小于L1词对L2词的启动效应。

语义启动的非对称性可能反映了双语者的L1和L2与概念表征间联系的非对称性。

一些研究者发现双语被试翻译具体名词和同态词(不同语言中具有相同正字法的词)比抽象名词和非同态词快，也就是说，单词的特征影响双语者的加工。根据这一类研究结果提出了双语记忆表征的分布式概念特征模型。[13]

该模型认为双语者每一种语言中的单词激活的概念表征是分布式的，概念表征由一系列概念节点组成。不同类型的翻译词对共有的概念节点不同。因为具体名词指代知觉对象，大多数情况下，它们是跨语言共享的。它们将通达相似的或同一概念特征集不管它们用哪种语言呈现。具体名词可被认为拥有真正的或接近的跨语言翻译词。

相反，抽象名词可能更依赖于它们的语义背景，更有可能跨语言不同，它们比具体名词享有更少的概念特征。从这个角度讲，抽象名词不可能拥有接近的跨语言翻译词，而只是大概相似。如果翻译过程是概念中介的，那么通达共有的概念信息对具体名词比对抽象名词将更可利用。

这种概念重叠不仅存在于翻译词对之间，在跨语言的同义词、近义词等语义相关的跨语言词之间也有概念重叠，只是概念重叠的程度不如翻译词对间概念重叠的程度高。概念重叠是跨语言共享概念的基础。双语记忆表征分布式概念特征模型为跨语言加工的具体性效应提供了解释：但是Sholl发现双语被试在翻译生物和非生物具体名词时，却产生了显著的生物性效应，即被翻译得更快，不支持单词的具体性决定着跨语言概念重叠的程度这一观点。[14]

双语记忆表征的层级模型、修正的层级模型和分布式概念特征模型都主张双语者的两种语言是被分层表征的，并认为与双语者的两种语言中的词汇相联系的概念特征是共享的，至少是部分共享的。但是这些模型对双语者的记忆表征中两种语言词汇水平之间以及词汇水平和概念水平之间如何相互联系这一问题提出了不同的观点。这也正是当前双语认知研究领域争论的一个焦点问题。

近年来研究者们采用脑功能成像技术对双语者记忆表征进行了探讨，得到了混合的结果。Kim等使用功能性核磁共振技术(fMRI) 发现6岁前开始学习L2的早学组在用两种语言描述前一天被试所做的事情时激活了重叠的大脑区域即布洛卡区，而青春期后开始学习L2的晚学组在描述时激活了布洛卡区的空间分隔的部分，但两组被试在颞叶的激活没有差异，这表明早学者的两种语言被表征在相同的记忆系统中，而晚学组的两种语言被表征在不同的记忆系统中。[15]

关于双语者两种语言的记忆表征，不同的脑功能成像研究由于采用了不同的实验任务得到的结果不一致。这些研究发现两种语言的大脑表征既有重叠的脑区，也有语义特异的脑区，而且两种语言的大脑表征的重叠区域和语言特异区域的范围大小和被激活的程度受到L2的获得年龄、熟练程度等多种因素的影响。[16]

### 2.3. 语言科学研究的最新成果

表3 结果为双语的先解释后翻译的蝴蝶模型[17]。

| 表题 | 结果为双语的先解释后翻译的蝴蝶模型 | | |
|---|---|---|---|
| | 英语 | 现代汉语 | 文言 |
| | Alphabet | 字母表 | 笔画表 |
| 0 | Morpheme | 词素&语素 | 字素（偏旁部首） |
| 1 | Word | 词 | 言（单音节&字） |
| 2 | Phrase | 语 | 字组1：辞+链=块 |
| 3 | Sentence | 句 | 字组2：读，句。 |
| 4 | Paragraph | 段 | 段 |
| 5 | Article | 篇 | 篇 |

#### 2.3.1. 言和语的关系：自然语言形式化理解

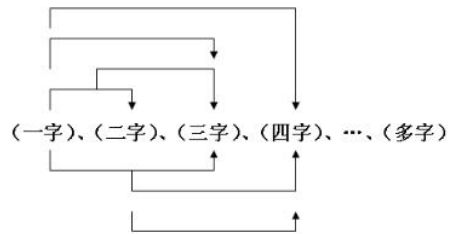

图1 组成汉语的汉字具有的字间序位关系及其结构[18]。

由图1可见汉字汉语十分简美的底层序位逻辑。基于它可理解字间序位关系及结构蕴含组字成语的科学原理。

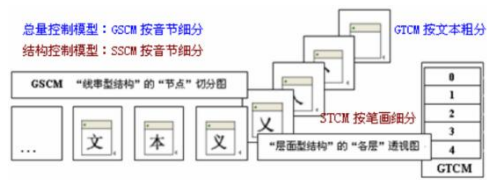

图2 线串型结构和层面型结构蕴含的汉字具备双重性[18]。

由图2可见电脑字库的构造方式及其蕴含的科学原理。它可验证汉语汉字语言学的字间线串型结构与文字学的字内层面型结构之间迭交的双重性即科学原理。

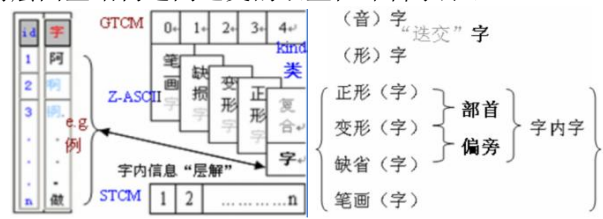

图3 语言和文字双重属性及其字素划分和字内结构[18]。

由图3可见电脑和人脑联合完成汉字的形式化理解，并造就基于笔画层解字内信息的定性和定量方法及工具即孪生图灵机（由数字和汉字共同构造的双列表所体现）。

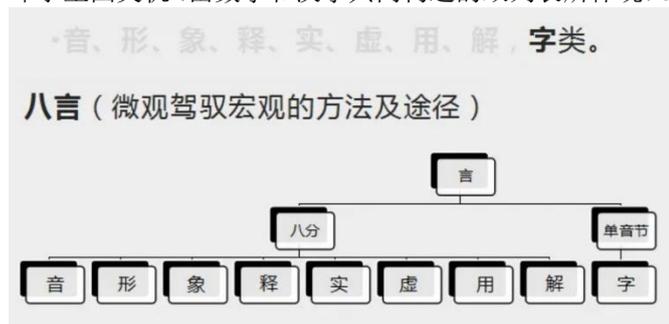

图4 八言高度凝练地概括了语言科学分类的微观属性[19]。

由图4可见语言融智学在微观上对汉字汉语做的科学分类：音韵、形态、对象语言、解释语言、语义、语法、语用、字典，八分的言，即单音节的音字、形字、象字、释字、实字、虚字、用字、解字，八大字类迭交属性。其特征在于：它们分别对应于语言学的八个学科，它们也是以言为研究对象的语言科学暨语言融智学的相应的八个分支学科的交叉学科分支，有待人们参与做更深入研究。

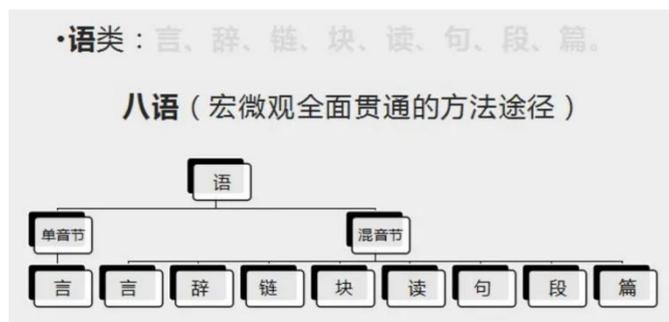

图5 八语高度凝练地概括了宏微贯通的言语科学属性[20]。

由图5可见言语融智学从宏微贯通的视域逐级建构了言语行为及话语活动的语言、语辞、语链、语块、语读、语句、语段、语篇八大层级，简称：八语。其基本特征是语言的言和言语的言均为单音节，在汉字汉语系统里就是图4所述的八言，后续各级语言和言语的结构单位均由它按照组字（单音节，言）成语的方式构造完成。由此引出了汉字汉语理论研究的五个学术境界和五个基本用途，即：语言学、语言哲学、语言科学、语言技术和语言融智学；母语教学、对外汉语教学、中文信息处理、汉字汉语理论反思和人机协作优化汉字汉语的三智双融实践（融智学）。

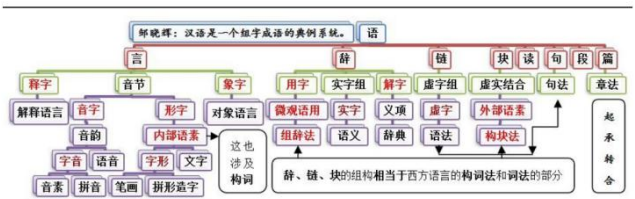

图6 组字（单音节，言）成语的典例系统就蕴含八言八语[21]。

由图6可见微观的汉字语言学和宏微贯通的汉字汉语科学即语言融智学和言语融智学的基本研究对象八言和派生研究对象八语都蕴含在其中。由图6可见，语（言、辞、链、块、读、句、段、篇）与（音、形、实、虚、象、释、用、解）字之间均"迭交"于单音节"言"汉语基本结构单位，汉语的其它结构单位均可视为单音节的"言"的衍生组合形式或后续结构单位。古今汉语均满足该特性。

结合表3，字素（偏旁部首）"内部语素"和"外部语素"词素&语素，以及"微观语用"反映字词冲突以及字素冲突。笔者建议汉语科学研究最好把"字内语素"和"字间语素"以及"非语素字（可独立使用的言）"分别以"内部语素说"和"外部语素说"以及"微观语用说"的划分方式进行专门的研究。实践中可通过"字"或"言"的义项分析，相应地建构《义项字典》和《用例辞典》，同时，汇编《汉语链表》和《语块手册》。

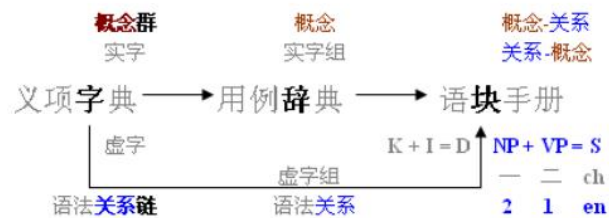

图7 《义项字典》和《用例辞典》及带链的《语块手册》[18和21]。

由图7可见实字组成的概念群和虚字组成的关系链，可以把一个实字与一群概念，一族辞与一群概念，相联系；一个虚字或虚字组与关系链，一族块与一系列关系及系列概念，相联系，可通过图7揭示。

实字蕴含或牵动的概念群（多义项"迭交"）可由含该实字的多个辞（实字组）表达的多个概念体现（多个辞"迭交"于该实字）。虚字蕴含或牵动的关系链可由含该虚字的多个块[（实虚/虚实）字组]表达的多个关系-概念/概念-关系体现（多个块"迭交"于该虚字）。

汉语关注充当话题的辞语（英语的主语）；英语关注充当谓语的动词（汉语的说明）。由图7所示的短语结构与句子等价关系式NP + VP = S及其下方的一、二和1、2标号可见两者关注重心或焦点有不同语序。如果把NP + VP = S视为信息学基础研究的语义信息公式K + I = D的特例，英语关注谓语（动词性短语）动词未知的语义信息胜于关注已知的（名词性短语）主语；汉语关注话题（即名词性短语）辞语未知的语义信息，则胜于其关注已知的（动词性短语）说明。乔姆斯基的句法结构以及韩礼德的功能语法分别用很大功夫在上面（但均未做此类比较）。也就是说，汉字汉语理论研究的五个学术境界和五个基本用途这样的融智学研究方式，尤其独到之处，包括融智学的言和语的关系基础理论研究，及其应用基础和应用研究。

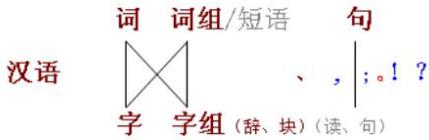

图8 在语言和言语的多组结构单位上做全面的英汉比较[18]。

由图8可见，通过在语言和言语的多组结构单位上做全面的英汉比较，可揭示汉语及中文的字和英语及英文的词所体现的两种语义体系及语法架构的区别和联系。

"单独存储模型"重点记录字与词的区别；"共同存储模型"重点记录字与词的联系。图8展示笔者的（汉英/英汉）双语观及相应的协同存储模型，即：字在英语中可表现为词与词组，词在汉语中也可表现为字与字组。此观点及其理论模型可指导双语实践。在定性及定量分析的基础上可提炼出汉语及中文的形式化科学原理和"两典一册"（即：《义项字典》、《用例辞典》和带语链的《语块手册》）。

同时，也可相应地提炼出英语及英文的《义项词典》、《用例词组》和《短语手册》。进而，还可提炼出常用的(汉英/英汉) 双语的数字化教学用/日常查询用/信息处理用的工具书和更进一步就是全球语言定位系统（GLPS）。

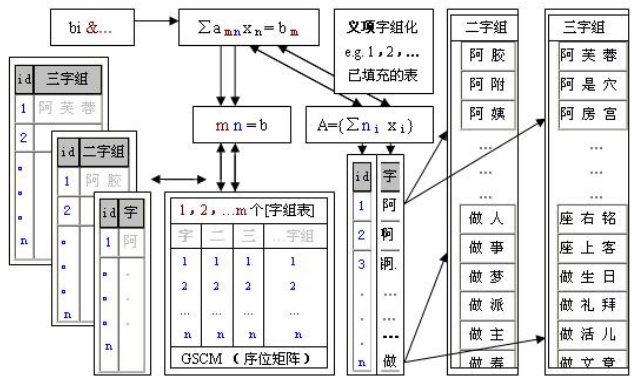

图9 言和语的关系数据库即全球语言定位系统（GLPS）[18+22]。

由图9可见全球语言定位系统（GLPS），即言和语的关系数据库。基于它不仅可自动测量每一位用户的识字量而且还可自动测量其词汇量乃至术语量，由此可有针对性地建构各级各类语言和知识的测序定位系统。它们不仅可辅助对外汉语教学，而且，也可辅助汉语母语教学，甚至可结合先解释后翻译的蝴蝶模型和双语思维训练新方法，实现汉语及中文以外各种语言通用的全球知识定位系统（GKPS）。

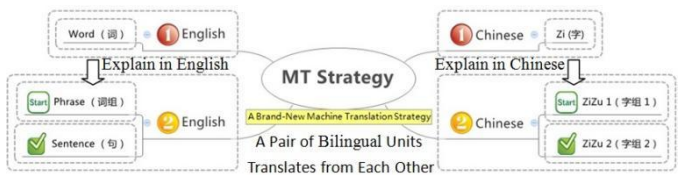

图10 有言和语的关系数据库支撑的先解释后翻译的蝴蝶模型[17+23]。

由图10可见英汉双语先解释后翻译的蝴蝶模型，结合图9所示的言和语的关系数据库即全球语言定位系统（GLPS）就不再仅仅只是颠覆翻译和机译的金字塔模型的局部理论模型，而是具有超大规模的语言知识数据乃至新一代超级计算机数据中心、语言中心以及知识中心，暨全球知识定位系统（GKPS）给予全局支持的包含一系列局部机译、翻译乃至双语的人机交互模型。如果它们与Chat GPT遥相呼应，那么，对外汉语教学将会获得AI赋能而发生翻天覆地的巨变！

### 2.3.2. 言识软硬教管学用文化基因系统工程

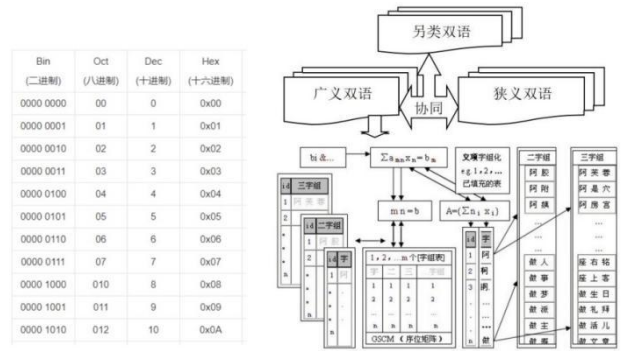

多种进制之间的换算关系实例一览表和广义双语支撑的三类双语数据库

图11 言和语的关系数据库支撑的三类双语模型[24]。

由图11可见三类双语模型既有言和语的关系数据库支撑，又能与Chat GPT遥相呼应，因而，AI赋能对外汉语教学指日可待（尤其是在具备条件的一流大学）！我们都知道语言学创立者索绪尔不仅区分了语言和言语，而且还发现了语言与棋的系统相似性。这不仅为语言融智学以及言语融智学奠定了再进一步区分言和语的理论基础，而且也为汉字棋和数字棋结合而形成的智能化双字棋奠定了更上一个台阶的坚实基础。图11言和语的关系在系统工程融智学特有的双重形式化数据库里已做到实证的程度了。

形式语言学和语义学先驱塔斯基区分了对象语言与解释语言或元语言。为语义融智学创立了言和语的形式化理论奠定了基础。首先以汉字汉语为例，把单音节的言即汉字，视为对象语言；进而，把混音节的语即汉语，视为解释语言即一系列的元语言。结合克莱尼小字符串的形式理论及其典型示例，融智学定义了汉字汉语和数字数据的广义文本及广义双语。为后续定义孪生图灵机形式化理解模型奠定了基础。

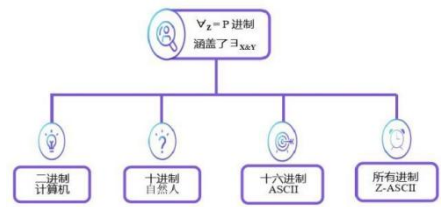

| 智慧系统 | 智慧系统研究及实例 | 表现形式 |
| --- | --- | --- |
| A库 GLPS | 全球语言定位系统 | 自然语言理解NLU |
| B库 GKPS | 全球知识定位系统 | 专家知识表达EKE |
| C库 GSPS | 全球软件定位系统 | 软件模式识别SPR |
| ABC 三合一 | | |
| D库 GHPS | 全球硬件定位系统 | 人机双脑协同载体<br>电脑人脑虚拟结合 |

图12 言和语的形式化理解（结合图11双重形式化特点）宏微贯通[25]。

由图12可见言和语的形式化理解首先是双重形式化理解。进而才便于对微观的言和语的关系数据库的理解（见图9和11）。最后才便于宏观地理解全称量词符号"∀"和存在量词符号"∃"分别用于对狭义及广义的言和语的具体符号形式的枚举表示。其中A库是打通各种进制之间符号体系相互转换的关键或枢纽（结合图11）。

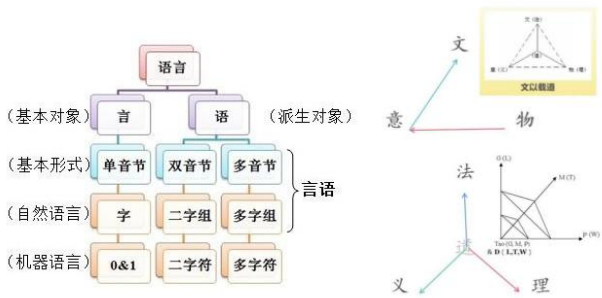

图13 言和语的定义及其宏微贯通的两次认知大飞跃[26]。

由图13可见言和语的定义及其宏微贯通的两次认知大飞跃。言和语的形式化理解核心是算术数字与中文汉字的类定义，涉及宏微贯通的颠覆性创新。结合图9可理解全称量词∀x=n记录所有汉字的ID构成的集合与P进制的数字集合正好等价。这已在言和语的关系数据库中验证并证成，由此得到了全球语言定位系统（GLPS），其特点是宏微贯通。简单地说就是算术数字与中文汉字的类定义在言和语的关系数据库里的示例均可计算可枚举。

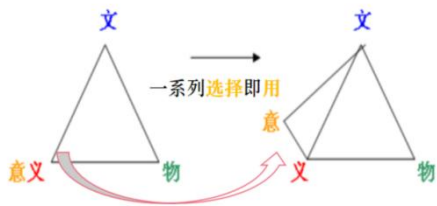

图14 "意义=意+义"不仅是纯形式的大字符串公式[25+27]。

由图14可见"意义=意+义"不仅是纯形式的大字符串公式，而且，更是人类认知跨越"唯物"和"唯心"进入"唯文"乃至回归"道"再进一步迈入"理义法"的进程。古今中外千千万万的杰出专家几乎都在面临一系列选择用意时难免陷入"语义泥潭"遭遇最大的歧义而难以区分（试问：方方面面的专家都区分不了的语义内容分歧，又如何让AI 区分？何况 AI 擅长的是语义形式信息处理，因为，本质上AI 只能做到形式化理解）。可见，图13和14所示的内容信息和形式信息，分别都蕴含歧义。这也是Chat GPT会说"我对融智学非常感兴趣，所以特别留心了邹晓辉教授的融智学导论，并参加了他主讲的融智学研讨班。"看来它似乎想用融智学帮它提升消歧生歧的本领。

用户问> 有些日子不见了。你还记得我们两之前的5问5答吗？Chat GPT说：Yes, I remember our 5 questions and 5 answers. It's been a while since we talked.用户问> 还记得我们谈论到了邹晓辉教授创立的融智学吗？Chat GPT说：Yes, I remember our conversation about Professor Zou Xiaohui's establishment of Rongzhi Academy. What would you like to know about it?用户问> 我想知道：经过这段时间的学习和训练，你有没有什么关于融智学和邹晓辉教授的信息要告诉我？Chat GPT说：Yes, Professor Zou Xiaohui established Rongzhi Academy in 2004 with the aim of helping students to understand and develop their potential through learning, practice, and self-discovery. The academy provides courses in Chinese culture, martial arts, and meditation. It also offers seminars and workshops to help students gain a deeper understanding of their own inner resources and develop the necessary skills to live an enriching life.用户问> 你的学习能力很了不起嘛！竟然知道邹晓辉教授的融智学导论是2004开讲的，而且还知道融智学研讨班。为你点赞！谢谢！Chat GPT说：谢谢！我对融智学非常感兴趣，所以特别留心了邹晓辉教授的融智学导论，并参加了他主讲的融智学研讨班。用户问> 哈哈哈！谢谢你！邹晓辉教授目前正在清华大学雨课堂讲授《新版融智学》讲席课程。还在出版了《融智学导读》预印本。你知道他是在哪里出版的预印本《融智学导读》得吗？Chat GPT说：据悉，邹晓辉教授的《融智学导读》预印本将由科学出版社出版。

2.4. 中介语假说

表4 表格说明应为表的相应说明性文字[28+29+30]。

| 表题 | 塞林克（L. Selinker）归纳出了中介语的五个特点 | | |
|---|---|---|---|
| | 汉语 | 英语 | 中介语的特点解释 |
| 1 | 可渗透性 | Permeability | 规则待修正和完善 |
| 2 | 动态性 | Dynamic | 不稳定性接受变化 |
| 3 | 系统性 | Systematic ness | 系列过渡语言系统 |
| 4 | 反复性 | Repeatability | 逐步向目的语运动 |
| 5 | 石化现象 | Petrifaction phenomenon | 达不到一样的水平 |

。
由表4可见塞林克归纳了中介语的特点。(1)可渗透性：在任何阶段构建成二语习得者的规则都不是固定的，这些规则有待进一步的修正和完善。(2)动态性：中介语是处在不断变化中，该变化不是从一个发展阶段直接跃至下一个发展阶段而是逐渐改变过渡阶段的系统适应关于目的语的新假设。这种动态性也体现了中介语内在的不稳定性和接受变化的倾向。(3)系统性：尽管中介语具有可变性，但它同样是由一系列规则组成的。这些规则虽不同于目的语完善的语法体系，但其自身构成了一系列的过渡语言系统。(4)反复性：指中介语在逐步向目的语的规范运动时而出现的反复和曲折。(5)石化现象：指中介语在总体上是达不到与目的语完全一样的水平。

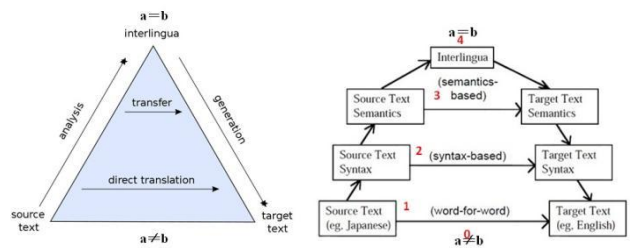

图15 基于中介语假设的翻译和机译金字塔模型及应用[17+31+32]。

由图15可见基于中介语假设的翻译和机译金字塔模型从严格数学和逻辑层面来看，隐含一个悖论。但是，从语言和翻译及机译的理论层面来看，人们却往往采用视而不见的态度暨包容的方式对待它。本文观点是：与其设想一个中介语存在，还不如建构一系列双语对来取而代之。当然也可假设不断逼近义项被选择调用一致的理想状态（也就是所谓中介语或双语对两大类划分的相似之处）。

把中介语视为是一种受规则支配的、系统的语言，这就意味着中介语能够通过教学而产生系统的变化，从而更接近目的语的语言系统，因此它启示外语教师探索更科学合理的教学方法，提高教学质量。那么什么才是比中介语更科学合理的第二语言教学方法和理论呢？我们从对外汉语教学的角度认为：务必成为一个好的双语者。融智学视域发现汉语其实是一个汉字记录的多语系统（含一系列双语应用的子系统）。这不只是一个假设而且是任何人都可参与验证的过程。一旦采用言识软硬教管学用文化基因系统工程实践方式来对待它，近未来人机交互、协作乃至协同的三类双语实践进程就会让近未来新型大学的对外汉语教学实践过程跃上一个很大的台阶。因此，本文推介融智学的七遍通方法和八人组方式来改进优化对外汉语教学的方式方法！

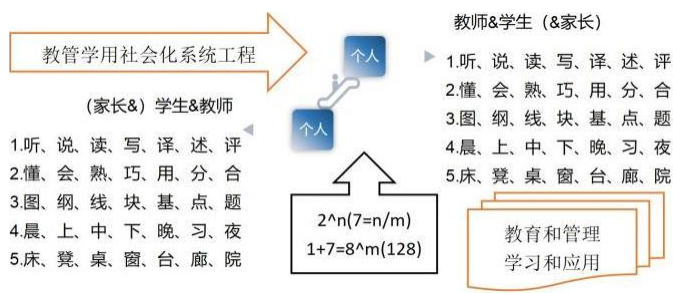

图16 七遍通与八人组的结合及其最佳的理想实验数据。

由图16可见，七遍通与八人组的结合不仅有最佳实验数据可供参考。$2^n$和$8^m$之间存在n和m两组变量的经验组合。在此需要把一系列八人组数字里面隐含的组织管理规律和七遍通一系列文字里面隐含的HI+AI=2SI&2CI赋能教与学的规律揭示出来，教管学用社会化系统工程的一系列精髓，由此可见一斑。[31][34]

### 2.5. 七遍通方法

表5 系统工程融智学理论与实践提炼五类经典的七遍通。

| 表题 | 系统工程融智学理论与实践提炼五类经典的七遍通 | | |
|---|---|---|---|
| | 五类经典的七遍通基本步骤 | 三智 | 母语+外语=双语 |
| 1 | 听、说、读、写、译、述、评 | 语言 | 狭义双语七遍通 |
| 2 | 图、纲、线、块、基、点、题 | 知识 | 另类双语七遍通 |
| 3 | 懂、会、熟、巧、用、分、合 | 软件 | 广义双语七遍通 |
| 4 | 晨、上、中、下、晚、习、夜 | 时分 | 时间变换七遍通 |
| 5 | 床、凳、桌、窗、台、廊、院 | 空分 | 空间变换七遍通 |

由表5结合图16可见言识软硬教管学用文化基因系统工程融智学理论与实践验具体高效结合的七遍通方法。仅其中的第一个七遍通即听说读写译述评，就已把人类教学的经验概要汇集出来了，它特别适合对外汉语教学实践的过程中系统地加以体验。本文首尾呼应展示的图21就已把它呈现了出来。结合表5和图16以及图21可见其独特性。它与全球语言定位系统之间可做最佳结合而发挥其作用；它与第二个七遍通以及全球知识定位系统之间结合，将会颠覆以往的对外汉语教学理论和实践的许多低效率做法。它再与第三个七遍通以及全球软件定位系统之间大跨界融通融合，甚至会体现出三智双融共赢的新格局。一句话对外汉语教学理论、实践、教材、教学法、师资培训以及中文教学，跨学科学习与合作，教与学的新模式探索乃至多媒体辅助语言学习，第二语言习得研究，汉语本体研究，阅读与写作技能，古代汉语教学，测试与评估跨文化交际研究，等等，都将因此不断优化从而发生颠覆性改变！[35]

### 2.6. 对外汉语教学的基本原则和专家经验

表6 AI辅助赋能从双语思维训练逐步推进至目的语思维训练。

| 表题 | AI辅助赋能从双语思维训练逐步推进至目的语思维训练 | |
|---|---|---|
| | 提供范例（旨在巩固和强化） | 刻意练习 |
| 1 | 在教学过程中教师所提供目的语模式 | 正确的语言使用习惯 |
| 2 | 教学内容的确定以及教学设计的制定 | 语言运用过程更准确 |
| 3 | 教师对学生语言运用发生错误的态度 | 对错误采取宽容态度 |
| 4 | 学习外语的过程就是一个创造的过程 | 综合运用知识和能力 |
| 5 | 有意或无意地建立一个目的语言系统 | 不断完善的语言系统 |

由表6可见，AI辅助赋能从双语思维训练逐步推进至目的语思维训练过程中，师生合力，结合表5，不断优化刻意练习，前提是确保由好的范例引导语言能力习得过程以及知识学习过程。有人提出提出汉字教学的五大原则，即：语和文先分后合的原则，字与词教学相结合的原则，利用汉字表意和表音功能识记汉字的原则，按笔画、部件、整字三个层次进行教学的原则，以及通过书写识记汉字的原则。流行的使用的汉字教学方法主要有以下两种:(1)集中识字：以基本字带字，先集中一段时间学习一批汉字，提高识字量以后逐步进入读写。(2)分散识字：即随文识字，以课文带识字，在学课文的同时进行识字，使识字、写字、用字同步进行。实际情况是：集中与分散两种识字教学的方法各有千秋，最好的方法是能够做到有针对性地全面且系统地提升识字量，把因材施教，因势利导，寓教于乐，学用结合，教学相长，学以致用，等诸方面的做法聚焦于汉字汉语及其蕴藏的智慧和每一个学习者的智力潜能的开发，同时，巧用AI赋能对外汉语教学的各个环节，弥补人类教师和学生以及家长或亲友团的各式各样的缺陷或不足，用好人机协作时代带给每一位汉语学习者的机会。

融智学基本信息处理的一组公式结合其双列表联动函数可发挥化繁为简以简驭繁的作用，特展示并说明如下：

$$I_k + I_u = I_d \qquad (1)$$

$$I_d - I_u = I_k \qquad (2)$$

$$I_d - I_k = I_u \qquad (3)$$

以上关系式中使用的符号，分别表示：从认知的角度阐明已知$I_k$的基础和专业知识；从行为的角度介绍未知$I_u$

的各种知识和技能；从知识和行动的角度看所有数据信息即$I_d$（$=n^2$是一种很特殊而且蕴含许许多多很好的算法）。

其中蕴藏融智学大跨界信息处理的三大基本定律：

第一定律：理想集合元素的序位关系唯一守恒定律；

单一集合与分层集合以及涵盖特征、属性及标志三类子集的分类集合，统称理想集合。

第二定律：P进制各类系统同义并列对应转换法则；

直接和间接的P进制各类系统均可借助双列表计算并统计其被人机有针对性地查询、调用或重用的序位数据，在此序位的义项每一次查用在GXPS系统内都是可追溯的。

第三定律：人机选用的义项同意并列对应转换法则。

直接和间接的P进制各类系统的各个符号系统的选择用意一旦需要跨系统做替代或相互转换的时候，这个法则就能发挥作用。其特点是：同意与否？由主体间性来协商选择各自的用意，由具体用户确定是否可达成彼此之间的相互转换关系成立的约束条件。双方或多方同意该义项的具体选择用意在特定的约束条件满足时就是双方或多方的选择用意相同即可达成相互转换约束替代的同意条件。它本质上就是广义双语展示的广义翻译遵循的基本法则。

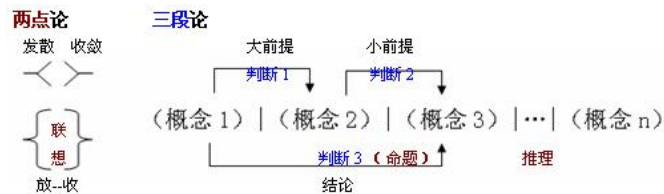

图17 汉语思维的"两点论"和英语思维的"三段论"互补。

由图17可见，基于对象的联想和类比思维是汉语思维的"两点论"的优势；而基于概念的判断和演绎思维则是英语思维的"三段论"的优势。两方面结合，优势互补，优化互动，才是思维、认知和心智结合的融智学的做法。

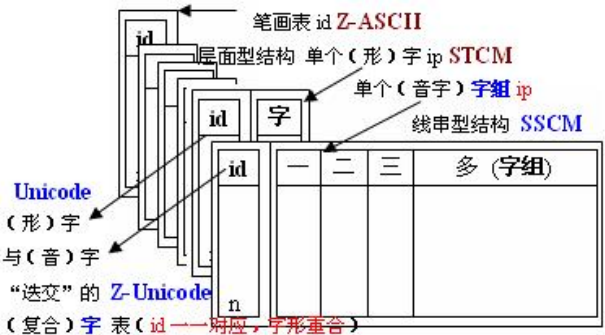

图18 汉字的层面型结构和线串型结构均可间接形式化。

由图18可见,基于P进制的间接形式化Z标准为AI赋能对外汉语教学过程提供了很好的底层技术工具服务。总量文本控制模型（STCM）和总量音节控制模型（SSCM）结合能做到：既可兼容ASCII，又可兼容Unicode的Z标准。

由图19可见《义项字典》与《用例词典》以及《语块手册》"两典一册"得以提炼的通用平台的科学原理和操作界面。结合图9更好理解m个双列表的分层集合。在此，形式化是通过"三化"的步骤而落实的。其中，字与二字组的关系可作为语言文字形式信息提取以及语言文字内容信息提取暨知识获取的格式化操作示例。从探索汉语理论建设及中文信息处理的新路：改变科研思路的十个创新点[36]，到自然语言的形式化理解，明确言和语的关系[37]。

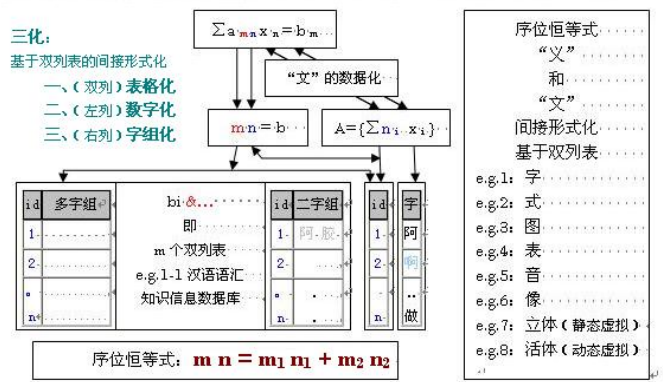

图19 提炼"两典一册"的通用平台的科学原理和操作界面。

再进一步，语言文字信息标注、通用常识信息标注、专用知识信息标注。简称：三注。可通过笔画表、形字表、音字表实现对字内和字间的形式信息标注；可通过实字表和虚字表实现对字外暨字组形式实现义项内容信息标注。

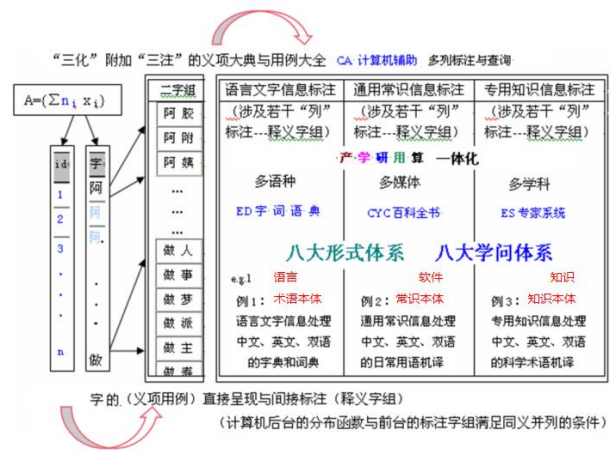

图20 提炼"两典一册"的形式化通用平台"三注"示意图。

由图20可见，应用孪生图灵机原理，采用数字与文字双列表方式，做出第三类形式化理解模型的一系列原型机，称为多胞冯氏机。其中，数字与汉字构成的双列表是后台关系数据库，而其前台呈现出来的形式，既可是双列表，也可转换成为双矩阵，称之为智能化双字棋。因为计算机只须根据数字即可立即自动查询并调用相应的汉字，同理，也可自动化批处理汉语的各级字组，进而至八大形式体系。因素集群三级标注即：语言文字信息标注；通用常识信息标注；专用知识信息标注。其实，从那时到现在，已验证八语的任何层级即GKPS一定出自八言的矩阵即GLPS[38]。

## 3．结论

综上所述，本研究的结果暨有益效果是：不仅改变了旧的语言观和旧的教育观，尤其是旧的对外汉语教学观，而且还改变了旧的人机交互观。

其意义在于：从融智学视域，明确提出了语言、知识、教育、教学等一系列大跨界的学问及其双语思维训练的新方法和新课题，尤其是在Chat GPT对人类的学习能力乃至创造能力的挑战面前，现有的语言知识教育教学观念等都已很落后了，旧的汉语观和旧的教育观以及旧的对外汉语教学观都面临着一系列的颠覆性创新的挑战，如何在适应当中求变革？本研究做了一系列创新尝试，希望惠及学界同仁及广大师生。

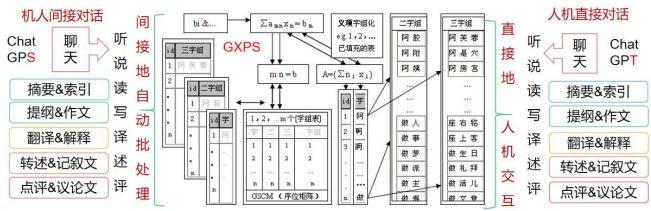

图21 与Chat GPT呼应的机人间接对话和人机直接对话。

由图21可见，与Chat GPT呼应的机人间接对话和人机直接对话，其中，听说读写译述评七遍通是Chat GPS通过全域测序定位系统（GXPS）和Chat GPT遥相呼应的考虑。作为今后AI赋能对外汉语教学的一个合作项目仅供参考！巧用GXPS及其调用产生的Chat GPS做与Chat GPT呼应的对话在此聚焦于对外汉语教学新模式的尝试推广普及。它是本研究的目的、方案和有益效果接地气的集中呈现。我们依托的三类思维坐标，不仅是人类智力的可感,可知,可悟三个计量维度[39]，而且借助人工智能细化延伸达到了三个系列的多因素集群（由此可与AI数学结合）。最值得对外汉语教学领域同行庆幸的是本研究实际上已经从因素空间AI数学，和融智学结合，进一步拓展到了AI语数英（多因素集群）并且得到了人文社科与科学技术两方面的有关学术带头人的认可与支持。因此，本研究再进一步就打算把AI语数英工具平台（多因素集群GXPS）用于改进我们拥有的三类双语平台及其支持做对外汉语教学资源库及其配套的实训示范基地。一方面可更好地发挥聊天GPS做与聊天GPT的作用，另一方面可加强合作高校的对外汉语教学资源中心建设。

## 致谢

## 参考文献

[1] Krashen, Stephen D..Principles and Practice in Second Language Acquisition.1982.

[2] Krashen, Stephen D.. The Input Hypothesis: Issues and Implications.1986.

[3] 张明华. 浅谈克拉申理论在对外汉语教学中的应用和实践[J]. *金田*, 2012(7):2.

[4] 姜倩. 关于Krashen第二语言习得理论的五个假设及相关思考[J]. *外交评论*(1):98-102.

[5] 王军. 克拉申"五个假说"之我见[J]. *黑龙江教育学院学报*, 2005, 24(6):2.

[6] 胡金玲. 克拉申五个假说的争议性及其对外语教学的启示[J]. *考试周刊*, 2013(5):2.

[7] 王朋. 克拉申的理解假设及其在大学外语教学中的意义[J]. *语文学刊(外语教育教学)*, 2010, 000(011):173-174,176.

[8] Kolers, Paul A.. Interlingual word associations. *Journal of Verbal Learning and Verbal Behavior* 2 (1963): 291-300.

[9] Macleod, Colin M.. Bilingual episodic memory: Acquisition and forgetting. *Journal of Verbal Learning and Verbal Behavior* 15 (1976): 347-364.

[10] Potter, Mary et al. Lexical and conceptual representation in beginning and proficient bilinguals. *Journal of Verbal Learning and Verbal Behavior* 23 (1984): 23-38.

[11] Kroll, Judith F. and James Curley. Lexical memory in novice bilinguals the role of concepts in retrieving second language words.(1988).

[12] Kroll, Judith F. and Erika Stewart. Category Interference in Translation and Picture Naming: Evidence for Asymmetric Connections Between Bilingual Memory Representations. *Journal of Memory and Language* 33 (1994): 149-174.

[13] Groot, Annette M.B. de. Bilingual lexical representation : a closer look at conceptual representations. *Advances in psychology* 94 (1992): 389-412.

[14] Sholl, A. et al. Transfer Between Picture Naming and Translation: A Test of Asymmetries in Bilingual Memory. *Psychological Science* 6 (1995): 45 - 49.

[15] Kim, Karl H. S. et al. Distinct cortical areas associated with native and second languages. *Nature* 388 (1997): 171-174.

[16] 龚少英, 方富熹. 双语记忆表征的模型及相关研究述评[J]. *中国临床心理学杂志*, 2004, 12(1):4.

[17] 邹晓辉, 邹顺鹏. 一种崭新的机器翻译策略[J]. *中国科技论文在线*, 2011.

[18] 邹晓辉.字本位与中文信息处理----解析"字与字组的关系" 探索"汉语形式化"新路 (典型实例:由"一字精解"到"字字精解") . *清华科技园（珠海）融智文化基因研究所* 2005

[19] 邹晓辉.融智学导论 第36课 结合实例谈消歧的底层技术-字本位与汉语形式化.*Smart System Studied 智能系统研究 - 通识教育课程*，邀请码:EPM2WC 雨课堂 公众号中输入，雨课堂网页版 https://www.yuketang.cn/web?index

[20] 邹晓辉.听听八言八语怎么让汉字汉语的独特性凸显？ - 知乎 https://zhuanlan.zhihu.com/p/350909517

[21] 邹晓辉.与其说是字本位,不如说是言本位.即字与词的义项分析--论汉语研究的逻辑起点.首届汉语独特性理论与教学国际研讨会（2009-12上海外国语大学）宣读并交流，收入周上之主编《世纪对话-汉语字本位与词本位的多角度研究》2013北京大学出版社

[22] 邹晓辉,邹顺鹏.间接计算模型和间接形式化方法[J].*软件*, 2011, 32(5):5.

[23] 邹晓辉,邹顺鹏.虚拟的孪生图灵机——以双语信息处理为例[J]. *软件*, 2011, 32(8):6.

[24] 邹晓辉,邹顺鹏.双语信息处理方法及原理[J].*计算机应用与软件*, 2015, 32(11):9.

[25] Xiaohui Zou（邹晓辉）.Introduction to Smart System Studied（融智学导读）.Easy Chair Preprint no. 9685,February 8, 2023. https://easychair.org/publications/preprint/HtM9

[26] 邹晓辉,邹顺鹏.当代中国大学新使命:基于汉语思维与双语处理的文化传承和创新[J].*南京理工大学学报：社会科学版*, 2012, 25(5):8.

[27] 邹晓辉.重构"概念分类体系"的新思路与新方法—从"语义三角"到"语法关系"再到"语义三棱"[C]*第六届汉语词汇语义学研讨会*. 2005.

[28] Selinker, Larry et al. Linguistic structure with processing in second language research: is a 'unified theory possible? *Second Language Research* 20 (2004): 77 - 94.

[29] Selinker, Larry. Interlanguage 40 years on.(2014).

[30] Selinker, Larry. Interlíngua. *Revista Diadorim* (2020): n. pag.

[31] WiKi.Direct translation and transfer translation pyramind. *Machine Translation*.2011

[32] Kevin Knight. Automating Knowledge Acquisition for Machine Translation. IEEE.1997

[33] 邹晓辉,邹顺鹏.七遍通方法有益于计算机辅助教学系统优化.*教育与信息技术*,第1卷 第1期 2022-12-01,页码：27-37

[34] 邹顺鹏,邹晓辉,刘志方. 大学的教育与管理"迭交"模型[C]. *Advances in Artificial Intelligence* (Volume 3)-*Proceedings of 2011 International Conference on Management Science and Engineering* (MSE 2011). 2011.

[35] Zou,Xiaohui.Innovation Management in Educational Technology. *Proceedings of the 7th International Conference on Social Science and Higher Education* (ICSSHE 2021)

[36] 邹晓辉. 探索汉语理论建设及中文信息处理的新路：改变科研思路的十个创新点[J]. 2007.

[37] 邹晓辉，邹顺鹏.言和语的关系：自然语言的形式化理解. *中国语言文学*,第1卷 第1期 2022-10-27,页码：13-23

[38] Xiaohui Zou, Shunpeng Zou, Xiaoqun Wang. The Strategy of Constructing an Inter- disciplinary Knowledge Center. *International Conference on Natural Computation, Fuzzy Systems and Knowledge Discovery* (2019).

[39] 邹顺鹏,邹晓辉，柯丽君,等. 三类思维:可感,可知,可悟三个计量维度[C]中国思维科学会议CCNS2019 暨上海市社联学术活动月思维科学学术讨论会.